\documentstyle{mn}
\input{epsf}
\newcommand{\ba}{\begin{eqnarray}}
\newcommand{\ea}{\end{eqnarray}}
\newcommand{\be}{\begin{equation}}
\newcommand{\ee}{\end{equation}}
\newcommand{\nn}{\nonumber \\}
\newcommand{\vk}{\vec{k}}
\newcommand{\vkone}{\vec{k}_1}
\newcommand{\vktwo}{\vec{k}_2}
\newcommand{\vkthree}{\vec{k}_3}
\newcommand{\kvone}{\vec{k}_1}
\newcommand{\kvtwo}{\vec{k}_2}

\title{Large-scale bias in the Universe II:  redshift space bispectrum}
\author[ L. Verde,  A.F. Heavens, S. Matarrese \& L. Moscardini]
{L. Verde$^{1}$, A.F. Heavens$^{1}$, S. Matarrese$^{2}$, L. Moscardini$^{3}$\\
$^{1}$ Institute for Astronomy, University of Edinburgh, Royal Observatory, 
Blackford Hill, Edinburgh EH9 3HJ, United Kingdom\\
$^{2}$ Dipartimento di Fisica {\em Galileo Galilei}, Universit\`{a} di
Padova, via Marzolo 8, I-35131 Padova, Italy\\
$^{3}$ Dipartimento di Astronomia, Universit\`{a} di Padova, vicolo
dell'Osservatorio 5, I-35122 Padova, Italy\\}

\begin{document}

\maketitle
\begin{abstract}
The determination of the density parameter $\Omega_0$ 
from the large-scale distribution 
of galaxies is one of the major goals of modern cosmology.  However,  
if galaxies are biased tracers of the underlying 
mass distribution, linear perturbation theory leads to a degeneracy 
between $\Omega_0$ and the linear 
bias parameter $b$, and the density parameter cannot be estimated. 
In Matarrese, Verde \& Heavens (1997) we developed a method based on
second-order perturbation theory to use the bispectrum to lift this 
degeneracy by measuring the bias parameter in an $\Omega_0$-independent way.
The formalism was developed assuming that one has perfect
information on the positions of galaxies in three dimensions.  In galaxy 
redshift surveys,  the three-dimensional information is imperfect, because 
of the contaminating effects of peculiar velocities, and the resulting 
clustering pattern in redshift space is distorted.
In this paper, we combine second-order perturbation theory with a model for 
collapsed, virialised structures, to extend the method to redshift space,
and demonstrate that the method should be successful in determining with
reasonable accuracy the bias parameter from state-of-the-art surveys such
as the Anglo-Australian 2 degree field survey and the Sloan digital sky survey.
\end{abstract}
 

\section{Introduction}

One of the most important challenges of observational cosmology today is 
to measure the density parameter $\Omega_0$, that plays a key role 
in the determination of the properties and the  geometry of the 
Universe. Geometrical tests such as number counts or the apparent magnitude 
as a function of the redshift of standard candles suffer from serious 
uncertainties, being sensitive to possible luminosity evolution with 
redshift and to the effect of a possible cosmological constant. This 
motivates attempting to measure $\Omega_0$ in the local Universe from 
large-scale structure (LSS).  Indeed, the only strong suggestions that
the Universe may be close to the critical density comes  
from peculiar velocities or redshift distortions studies of LSS. 
These studies have the advantage that evolution and geometrical 
effects are negligible, but have some drawbacks.
The amplitude of the distortions on linear scales allows 
measurement of the density parameter only in the 
degenerate combination
$\beta = \Omega_0^{0.6}/b$ where $b$ is the linear bias parameter 
(e.g. Kaiser 1987; Hamilton 1992; Fisher, Scharf 
\& Lahav 1994; Heavens \& Taylor 1995; Ballinger, Heavens \& Taylor 
1995; Cole, Fisher \& Weinberg 1995).
The current values for $\beta$ fall roughly in the range 0.5-1 depending on 
the measurement method used and on the catalogue selection criteria 
(Strauss \& Willick 1995; Hamilton 1997).  Even if $\beta = 0.5$, 
$\Omega_0$ could still be close to its critical value if dark matter is 
poorly traced by luminous objects.  The determination of the density
 parameter from linear peculiar velocities or redshift space distortion
studies is therefore compromised by the lack of a good measurement 
of the bias parameter. This motivates methods which attempt to measure
$\Omega_0$ directly (Dekel \& Rees 1994; Fan, Bahcall \& Cen 
1997) or methods to determine $b$, which then allows determination of
$\Omega_0$ from $\beta$. In this paper, we pursue the latter approach: there
is a relative bias between IRAS and optically selected galaxies (e.g. Peacock
1997), so at least one of the galaxy types must be at best a biased tracer of 
the underlying mass distribution.
In practice, we have very little idea of how bias works and evolves, and 
there is therefore  strong motivation to measure it 
empirically from the galaxy distribution.  
Future Cosmic Microwave Background experiments may allow the 
simultaneous determination of all the main cosmological parameters 
with unprecedented accuracies (Jungman et al. 1996; Bersanelli et al. 1996),
in which case the major motivation of studies like this will be to constrain
galaxy formation mechanisms and the power spectrum by measuring $b$ directly,
rather than, as here, on using it as a route to $\Omega_0$.

In a previous paper (Matarrese et al. 1997; hereafter MVH97) we 
presented a method to measure the bias parameter from LSS data 
by studying higher-order characteristics of the density field. 
We developed an idea of Fry (1994): since the degeneracy between 
$b $ and $\Omega_0$ is an intrinsic feature of linear theory, 
one needs to go to second order to separate the parameters. 
Under the assumption that the initial fluctuation field is Gaussian 
and that structures grow by gravitational instability, the 
three-point correlation function and its counterpart in Fourier 
space, the bispectrum, are intrinsically second order quantities.
The bispectrum has the advantage that the range of validity of
perturbation theory appears well-defined 
(Scoccimarro et al. 1998) - if the power spectrum is roughly known - 
and in the appropriate range, the correlation properties  
of the estimators can be calculated. If the bias is local, 
this allows us to estimate it via a likelihood method, and therefore to 
assign an errorbar.
For consistency with second order perturbation theory (2OPT) the first 
non-linear bias term must also be considered, and this can introduce 
an additional
degeneracy between skewness induced by gravitational evolution and that 
introduced by bias.
The two effects can, however, be separated by the use of shape information: 
essentially non-linear biasing of a truly Gaussian field will lead to 
different shaped structures from a non-Gaussian field arising from
gravitational instability.
Since the signal comes from the mildly non-linear regime, the bias 
can be measured on scales where $\Delta^2$(the conventional variance
 per unit logarithmic interval) is about unity.  
In MVH97 we showed that the bispectrum method we presented 
succeeds in recovering the true value for the  bias  with a 
reasonable error in a very idealized case where
the positions of the particles were known in  real space.
In reality, the best distance indicator for large galaxy catalogues is 
redshift;  the main purpose of this paper is to show how the 
bispectrum in redshift-space can be used to estimate bias.

Galaxy catalogues use the redshift as third spatial coordinate.
In a perfectly homogeneous Friedman Universe, redshift would be an 
accurate 
distance indicator, but inhomogeneities perturb the Hubble flow,
and introduce peculiar velocities.
The resulting redshift-space map of the galaxy 
distribution is thus distorted.
As Kaiser (1987) pointed out, peculiar velocities distort the clustering 
pattern in redshift space on all scales, but the effect can be regarded 
as being split into two components: a large-scale distortion resulting 
from coherent inflow into over-dense 
regions, and the `Fingers-of-God' (FoG) arising from virialised, highly 
non-linear structures.
For the bispectrum, we find distortions in redshift-space arising in 2OPT
(cf. Hivon et al. 1994), but also significant effects from virialised
structures. 
In this paper, we incorporate the effects of virialised structures by
a method which has been successfully used with the redshift-space power
spectrum: we model the effects as an incoherent velocity dispersion.
Once again the shape information allows the disentanglement of redshift space 
distortions and other effects:
redshift-space mapping modifies the shape of the structures in a 
different way from gravitational evolution and biasing.

The next generation of galaxy surveys, like for example the Anglo-Australian
2-degree Field (2dF, Colless 1996) and Sloan Digital Sky Survey 
(SDSS, Gunn \& Weinberg 
1995), will allow not only accurate measurement of $\beta$, but also 
estimation of the bias parameter with an accuracy of a few percent. 
 
This paper is principally concerned with the treatment of redshift 
distortions; a subsequent paper will present a detailed study of the other 
issues.

The plan of this paper is as follows. In Section 2 we quantify the effect 
that redshift distortions have on the bispectrum, and, as a consequence, 
on the determination of the bias parameter. Then we present a second-order 
perturbation description for the effect of redshift distortions on the 
bispectrum and 
on the covariance matrix needed for the likelihood analysis.
Also we discuss how to include the modelling of small scale, 
highly non-linear effects, such as FoG:
non-linear dynamics and virialised structures in redshift space 
contaminates scales that in real space would be only mildly non-linear.  
Section 3 contains the practical implementation of the method in a 
numerical simulation of the redshift space galaxy distribution.
In Section 4  we discuss the applicability of the method in more general 
terms: we implement the analysis on a open Universe 
simulation and on a
biased distribution in real space. The results obtained from the unbiased and
the biased distributions are also shown. 
Section 5 discusses the results also in the prospect of the next generation 
of large galaxy redshift surveys.

\section{Modelling of redshift distortions}

It is worth making a few remarks about performing this sort of analysis in 
Fourier space. Fourier space analysis has the great advantage that there 
is a clear separation of scales where perturbation theory works and breaks 
down.
Given the nature of the redshift distortion however, the natural way would
 be to decompose the density field in spherical harmonics and spherical 
Bessel functions (Heavens \& Taylor 1995). But, we showed in MVH97, rather 
than using the survey as a whole, it is more effective to split it into 
subsamples. 
For a deep survey in the individual subsamples the Fourier decomposition 
and Kaiser's (1987) distant observer approximation to model even large-scale 
distortions is a good description. 

\subsection{Effects of redshift distortions on the power spectrum and 
on the bispectrum}

As a preliminary, in this section we compute the effects of redshift
distortion on a particularly simple class of bispectra, those whose wavevectors
form an equilateral triangle and averaging over orientation.  This 
analysis is not required in later sections, but serves to illustrate
the effects of perturbation theory and the virialisation model.

On large (linear) scales the redshift space effect on an individual Fourier 
component of the density fluctuation field $\delta_{\vk}$ can be described as:

\be
\delta_{\vk} \longrightarrow \delta^s_{\vk} = \delta_{\vk} (1 + \beta \mu ^2)
\;,
\label{basic}
\ee
where the superscript $s$ refers to the quantity in redshift space, and $\mu$ 
is the 
cosine of the angle between the $k$-vector and the line of sight (i.e. the 
direction along which the distortion takes place).  Throughout, we adopt the
distant-observer approximation, so $\mu$ is independent of location within
the survey.

The effect on the average power spectrum in a  thin shell in $k$-space is:
\be
\langle\delta_{\vk}\delta_{\vk^*}\rangle \longrightarrow 
(1+\frac{2}{3}\beta+\frac{1}{5} \beta^2)
\langle\delta_{\vk}\delta_{\vk^*}\rangle \;.
\ee
This is an enhancement of the redshift space power on all scales by the same 
factor, 28/15 in the case where $\beta=1$.

The bispectrum $B(\vec{k}_1,\vec{k}_2,\vec{k}_3) 
\propto \langle \delta_{g \vec{k}_1} \delta_{g \vec{k}_2} 
\delta_{g \vec{k}_3}\rangle$ is non-zero only if the 3 wavevectors form a 
closed triangle.  For equilateral triangles, averaging the three factors in 
equation (\ref{basic}) over $\mu$ gives
\be
B^s=(1+\beta+\frac{3}{5}\beta^2+\frac{1}{7}\beta^3)B \;,
\ee
so that for $\beta$=1 there is an amplification of the bispectrum on all 
scales by a factor $B^s/B=96/35$.

We can estimate the effect of redshift distortions on the estimation of the
bias parameter by noting that, in the absence of shot noise and a non-linear 
bias term, $b \propto P^2/B$ (MVH97). If redshift distortions are ignored,
the bias parameter is overestimated by a factor 
\be 
\frac{(P^s/P)^2}{B^s/B}=1.27 \;,
\label{ratio} 
\ee
if $\beta=1$.

Virialised motion produces a radial smearing and the associated FoG 
effect contaminates the wavelengths we are interested in. 
This is hard to treat exactly, but being a smearing effect it produces 
a mild damping on the power acting in the opposite direction
to the large-scale boosting of power 
(see for example Matsubara 1994). A model was  
introduced by Peacock (1992; see also Peacock \& Dodds 1994)  in which 
the  small scale velocity field is assumed to have an incoherent Gaussian 
distribution.
In reality, according to evidence from simulations (e.g. Zurek et al. 
1994) and observations (e.g. Marzke et al. 1995) the velocity distribution 
is better modelled by an exponential, but in practice there is very 
little difference between these two models on scale where the damping 
factor is $\leq 2$ (Ballinger, Peacock \& Heavens 1996). In practice this incoherent model
fits well the power spectrum in numerical simulations (Hatton \& Cole 1998). 

In the Fourier domain the exponential velocity dispersion model gives a 
damping factor $D(k\sigma\mu)$ (where $k\equiv|\vk|$), such that
$\delta_{\vk}\longrightarrow\delta_{\vk}D$, given by
\be
D(k\sigma\mu)=\frac{1}{\sqrt{1+k^2\sigma^2\mu^2/2}} \;,
\label{damping}
\ee
where $\sigma$ is the pairwise velocity dispersion of galaxies.
The overall effect in $k$-space can be obtained by multiplying the large-scale 
enhancing factor and the small-scale damping factor before 
averaging over $\mu$ (Peacock \& Dodds 1994):
\be
\delta_{\vk}\longrightarrow\delta_{\vk} \frac{(1+\beta \mu^2)}
{\sqrt{1+k^2\sigma^2\mu^2/2}} \;.
\label{PD}
\ee
For the power spectrum in a thin shell in $k$-space the overall 
effect is given by:
\ba 
& & P^s(k)=   \nn
& & \left\{4\frac{(\sigma^2k^2-\beta)\beta}{ \sigma^4k^4}+\frac{ 2\beta^2 }
{3 \sigma^2k^2}+\right. \nn
& & \left.\frac{\sqrt{2}(k^2\sigma^2-2\beta)^2 \arctan(k\sigma/\sqrt{2})}
{k^5\sigma^5}\right\}P(k) \;. \nn 
\ea

Similarly for the angle-averaged bispectrum for equilateral triangles one 
obtains
\ba
& & B^s(k)= \nn
& &\left\{\frac{2k^6\sigma^6 - 12 k^4\sigma^4\beta+(36k^2\sigma^2+6k^4
\sigma^4)\beta^2}{\sqrt{2}k^6\sigma^6 \sqrt{2+k^2\sigma^2}}\right.\nn
& &+\frac{(k^4\sigma^4-5k^2s^2-30)\beta^3}{\sqrt{2}k^6\sigma^6 \sqrt{
2+k^2\sigma^2}}+  \nn
& & \left.\frac{3\sqrt{2}\beta(2k^4 \sigma^4-6k^2\sigma^2\beta+5\beta^2) 
\mbox{ arcsinh}(k\sigma/\sqrt{2})}{k^7\sigma^7}\right\}B(k) \;. \nn
\ea

\begin{figure}
\begin{center}
\setlength{\unitlength}{1mm}
\begin{picture}(90,70)
\includegraphics{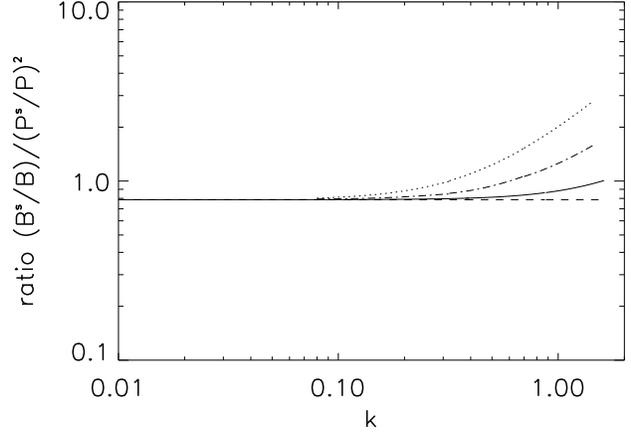}
\label{figratio}
\end{picture}
\end{center}
\caption{Scale dependence of the redshift distortions on the measurement 
of the bias parameter, which is overestimated by 1/ordinate if the distortions
are ignored. The dashed line refers to the case where there is 
no small scale velocity dispersion ($\sigma=0$),  the solid line is for 
$\sigma=200$ km s$^{-1}$, the dot-dashed line is for $\sigma=400$ km s$^{-1}$
and the dotted line is for $\sigma=700$ km s$^{-1}$. The wavenumber $k$ 
is in units of $h$ Mpc$^{-1}$.}
\end{figure}

\begin{figure}
\begin{center}
\setlength{\unitlength}{1mm}
\begin{picture}(80,70)
\includegraphics{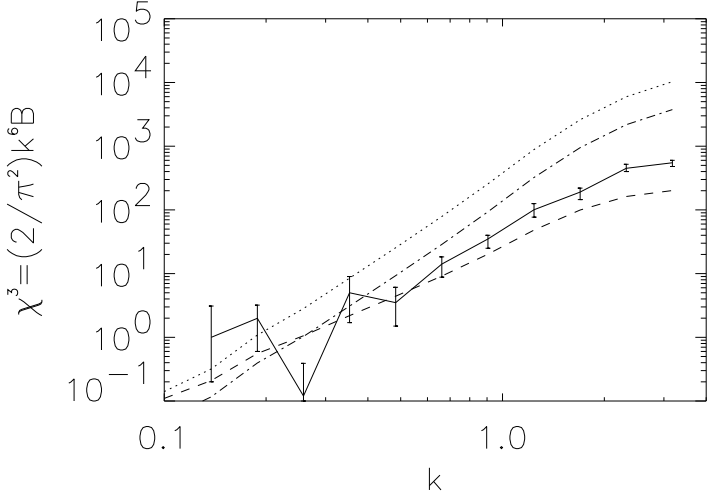}
\label{figbisp}
\end{picture}
\end{center}
\caption{Redshift distortions effects on the bispectrum. The bispectrum 
is expressed in the dimensionless form $\chi^3$ that is the counterpart 
for the bispectrum of  $\Delta^2$ for the power spectrum:
$\chi^3 \equiv (2/\pi^2)^2 k^6 B$, where $k$ is the smallest  wavenumber in
the triangle. The dot-dashed line is the 2OPT-real space bispectrum, the 
dotted line is the redshift-space bispectrum obtained taking into account 
only the large scale boosting effect, the  dashed line 
is the 2OPT bispectrum with the small scale smoothing effect included 
as in equation (8). The velocity dispersion used here 
($\sigma = 650$ km s$^{-1}$) is the one that gives the best fit 
to the redshift-space power spectrum up to $k \simeq 0.8$. Finally the
thick solid line with errorbars is the measured 
bispectrum from the redshift-space map. The agreement up to $k\simeq 0.7$
should be compared with the corresponding breakdown of 2OPT in real-space
($k=0.55$; MVH97). As before $k$ is in units of $h$ Mpc$^{-1}$.}
\end{figure}

The overall effect on the measurement of the bias parameter is given by 
the ratio [equation (4)]. In the 
case $\beta=1$ (that, since $\beta$ is expected to be $\le 1$, is the 
case where large-scale redshift distortions are most important) the 
scale dependence of this quantity  is shown in Fig. 1. For small 
velocity dispersions  the effect is small for a reasonable 
range of $k$-vectors, but for bigger  $\sigma$ it diverges quite quickly. 
Since we expect to use wavelengths up to $k\simeq 2$ 
(throughout the $k$ unit is  $h$ Mpc$^{-1}$)
in order to have a good signal to noise, ignoring redshift distortions 
introduces unacceptably big errors.
In the next section we illustrate a more elegant way to deal with this 
problem and to have a more satisfactory result: the redshift distortions 
should be modelled for consistency to second order using perturbation theory
and the small scale effect cannot be neglected.

The redshift-space effect on the bispectrum from equilateral triangles is 
illustrated in Fig. 2, which shows results from an N-body simulation
by the Hydra consortium (Couchman et al. 1995). 
It is a cold dark matter (CDM) simulation with 
$\Omega_0=1$, a shape parameter $\Gamma=0.25$, $\sigma_8=0.64$, no cosmological constant and no biasing.

\subsection{Second-order perturbation description for the 
effect of redshift distortions}

In the framework of second order perturbation theory (2OPT) for the 
density field, for consistency also redshift distortions should be 
treated to second order.  Unfortunately highly non-linear effects 
contaminate scales where 2OPT should hold. This can be easily 
understood by considering how a spherical over-density appears 
distorted by peculiar velocities when observed in redshift 
space (in the small angle approximation).
At large scales (linear regime) the collapsing shells appear 
squashed, the squashing increases until the infall velocity 
cancels with the Hubble expansion. At this point, where the 
dynamics is only mildly non-linear, the region appears to be 
collapsed on a sheet of infinite density in redshift space. At 
smaller scales the collapsing regions and the already virialised 
(highly non-linear) regions appear elongated along the line of sight (FoG). 
>From equation (\ref{PD}) it is clear that the damping factor acts as a filter 
on the Fourier components of the density field. Therefore our analysis 
involves 2OPT of the density field and of the redshift distortions 
combined with the exponential model for small scale velocity dispersions 
acting as a smoothing filter.

\subsection{Large-scale second order model}

We assume a local correspondence between the galaxy overdensity field
$\delta_g$ and the underlying mass density field $\delta \equiv \delta\rho/
\rho$, and make a Taylor expansion in $\delta$:
\be
\delta_g = \sum_{i} \frac{b_i}{i!} \delta^i \;,
\label{taylorexp}
\ee
retaining terms to $i=2$, but ignoring $i=0$ as it contributes only to
$\vec{k}={\bf 0}$.

The bispectrum $B$ is defined by the 3-point function in Fourier space:
\be
\langle\delta_{g \vec{k}_1} \delta_{g \vec{k}_2} 
\delta_{g \vec{k}_3}\rangle =(2\pi)^3 B(\vec{k}_1,\vec{k}_2,\vec{k}_3)\delta^{\rm D} 
(\vec{k}_1 +\vec{k}_2 +\vec{k}_3) \;,
\ee
where $\delta^{\rm D}$ is a 3D Dirac delta function. The expression for
$B$ is given in MVH97 for
$\delta$ in real space. In redshift space, the bispectrum is inevitably
more complicated (Heavens et al., in preparation):
\ba
& & B(\vec{k}_1,\vec{k}_2,\vec{k}_3) = \nn
& & b_1^2 (1+\mu_1^2\beta)(1+\mu_2^2\beta) \nn
& & \left\{\left[2 P(\vec{k}_1) P(\vec{k}_2) 
{\rm Ker}(\vec{k}_1,\vec{k}_2,\beta,\mu_1,\mu_2,\mu)+cyc.\right]\right.  \nn
& & +\left. b_2 \left[ P(\vec{k}_1) P(\vec{k}_2) + {cyc.} \right] 
\right\} \;,
\label{bisp}
\ea
where the cyclic terms are (1,3) and (2,3), and
\ba
\vk_{3} & = & -\vkone-\vktwo \;, \nn
\mu_i & = & \frac{\vec{r}\cdot \vec{k}_i}{rk_i} \;, \nn
\mu  & = & - \mu_3 \;. \nn
\ea
The kernel is given by
\ba 
& & {\rm Ker}(\vec{k}_1,\vec{k}_2,\beta,\mu_1,\mu_2,\mu)\equiv\nn
& & J(\kvone,\kvtwo)b_1+\mu^2\beta b_1 K^{(2)}(\kvone,\kvtwo)+\mu_1^2
\mu_2^2\beta^2 b_1^2+\nn
& & \frac{b_1^2\beta}{2}(\mu_1^2+\mu_2^2)+\frac{b_1^2\beta}{2}\mu_1
\mu_2(\frac{k_1}{k_2}+\frac{k_2}{k_1})+\nn
& & \frac{b_1^2\beta ^2}{2}\mu_1\mu_2(\mu_2^2\frac{k_2}{k_1}+\mu_1^2
\frac{k_1}{k_2}) \;,
\label{kernel}
\ea
where the real-space kernel is (e.g. Catelan \& Moscardini 1994a)
\ba
& &J(\vec{k_1},\vec{k_2})\equiv  \nn
& &\frac{5}{7}+ \frac{ \vec{k_1}\cdot \vec{k_2} } 
{ 2 k_1k_2 } \left( \frac{k_1}{k_2} + \frac{k_2}{k_1} \right)+\frac{2}{7}
\left( \frac{ \vec{k_1}\cdot \vec{k_2}}{k_1k_2} \right) ^2
\ea
and $ K^{(2)} $ comes from the velocity field (e.g. Catelan \& 
Moscardini 1994b)
\ba 
& & K^{(2)}(\kvone,\kvtwo)=\nn
& &\frac{3}{7}+ \frac{ \vec{k_1}\cdot \vec{k_2} } 
{ 2 k_1k_2 } \left( \frac{k_1}{k_2} + \frac{k_2}{k_1} \right)+\frac{4}{7}
\left( \frac{ \vec{k_1}\cdot \vec{k_2}}{k_1k_2} \right) ^2 \;.
\ea

In real space the 2OPT bispectrum is very weakly dependent on $\Omega_0$
(e.g. MVH97).  In redshift space, there is a strong $\Omega_0$-dependence,
but only via $\beta$, which can be measured from the power spectrum.
The residual dependence on $\Omega_0$ is again very weak, and we therefore
quote the results for $\Omega_0=1$.

We wish to write $B$ in terms of observable quantities, which $P$ is not. To
the order which we are working, we may write $P=P_g/b_1^2$. There are one-loop
corrections to this which may be significant if $b_2$ and $b_3$ are not close
to zero (Heavens et al. in preparation). In Section 4 
we investigate the practical
effects of those corrections using biased simulations, but consistently with
perturbation theory, in terms of the real-space galaxy power spectrum $P_g =
b_1^2 P$, we can write:
\be
\langle\delta_{g \vec{k}_1} \delta_{g \vec{k}_2} 
\delta_{g \vec{k}_3}\rangle^s ={\cal G}(P_g,\beta,c_1,c_2,\kvone,\kvtwo,
\vkthree) \;,
\label{K2}
\ee
where
\ba
& & {\cal G}(P_g,\beta,c_1,c_2,\kvone,\kvtwo,\vkthree)=\nn
& & (2\pi)^3(1+\beta\mu_1^2)(1+\beta\mu_2^2)\nn
& & \left\{c_1 \left[2 P_g(\vec{k}_1) P_g(\vec{k}_2) {\rm Ker}
(\vec{k}_1,\vec{k}_2,
\beta,\mu_1,\mu_2,\mu)+ cyc.\right] \right. \nn
& &\left. + c_2  \left[ P_g(\vec{k}_1) P_g(\vec{k}_2) + {cyc.} 
\right] \right\}\delta^{\rm D} (\vec{k}_1+\vec{k}_2+\vec{k}_3) \;,\nn
\label{G}
\ea
where the parameters we wish to extract $b_1$, $b_2$ appear in the 
combinations
\be
c_1 \equiv {1\over b_1};\qquad c_2 \equiv {b_2\over b_1^2} \;.
\ee
It is interesting to notice here that the degeneracy between $\Omega_0$ 
and $b$ can be lifted because equation (\ref{kernel}), as explained
before, is dependent on $\Omega_0$ mostly through the measurable parameter
$\beta$ and is quite insensitive 
to $\Lambda$ (see also Bernardeau 1994; Eisenstein 1997);
the error introduced  by neglecting the $\Omega$-- and $\Lambda$--dependence
of the bispectrum is much smaller that the final expected error in the
determination of the bias parameter, and can therefore be safely ignored
for this purpose.
 Redshift-space distortions may also be introduced by a
separate effect,
arising from an incorrect geometry being used to create the redshift-space map
(Phillips 1994; Ballinger et al. 1996). This effect should be detectable with
SDSS and corrected for.
This accounts for the large-scale effects via perturbation theory.  We
now turn to the effects of small-scale velocities.

\subsection{Small scale model}

It is worth noticing that the model of small-scale velocity dispersion 
is simplistic because it takes no account of the fact that the velocity 
dispersion is correlated with the density field. This means that the 
dispersion is higher in high density regions such as galaxy clusters. 
The value $\sigma$ used in the model (equation 6) is therefore only an 
``effective'' velocity dispersion which depends on how galaxies 
populate the clusters and on the bias parameter $b$ (Fisher 1995).
In principle, the form of the filtering may be unconnected with the
small-scale velocity distribution, because the modelling assumes that the
velocity is uncorrelated with the density, which is untrue in detail.
In practice, then, one should perhaps treat the filter function as
a heuristic object whose parameter $\sigma$ is determined from the
large-scale data, by fitting the power spectrum.   A Taylor expansion
as $k\rightarrow 0$ will demand a quadratic form for 
$D \simeq 1 - k^2\sigma^2\mu^2/4 + O(k^4)$, so one might expect the 
model to be generally good provided $ k^2\sigma^2\mu^2$ is small. 
We have found this to be the case for the power spectrum;  in high-$\Omega_0$
models, imposing a limit of $\alpha=k^2\mu^2=0.3$ works well (for a CDM
model with
$\Omega_0=1$, $\sigma_8=0.64$ and $\Gamma =0.25$).

The small scale damping effect acts as a smoothing filter (see equation 
\ref{PD}) on the non-linear field.
Following the general mathematical method outlined in Section 3 of 
MVH97 for calculating $N$-point distributions in Fourier space it is 
possible to obtain an expression for the galaxy N-point spectra  in redshift
space that includes the large scale effect to second order in perturbation
theory (equation \ref{bisp}) and the small-scale effect due to the velocity 
dispersion of the galaxies (equation \ref{damping}). 
The technical derivation can be found in Appendix A.
For $N=3$, as a special case we  obtain the final expression for the galaxy 
bispectrum in redshift space for a 
given triangle configuration, that also includes small scale velocity 
dispersion:
\ba
& & \langle\delta_{g\kvone}\delta_{g\kvtwo}\delta_{g\vkthree}\rangle^s= \nn
& & \frac{{\cal G}(P_g,\beta,c_1,c_2,\kvone,\kvtwo,\vkthree)}
{\sqrt{\left(1+\kvone^2\sigma^2\mu_1^2/2\right)
\left(1+\kvtwo^2\sigma^2\mu_2^2/2\right)\left(1+\vkthree^2
\sigma^2\mu_3^2/2\right)}} \;,\nn
\label{bisps}
\ea
where ${\cal G}$ is given by equation (\ref{G}).

This is also the key equation of this paper,  generalising MVH97 to 
redshift space galaxy catalogues.   Note that we assume that $\beta$ 
and $\sigma$ (and hence $P_g(k)$ in real space) are computed directly
from the redshift-space power spectrum $[P_g^s(\vec{k})]$, so they are
not parameters to be determined from the bispectrum (see section 3.1).

In order to be able to perform the likelihood analysis the 
covariance matrix for the bispectrum has to be consistently modified.
The covariance matrix for the bispectrum involves the  expression for 
the six-point function as shown in equations (15) and (16) of MVH97.

Referring to equations (38) to (42) of MVH97, we will quote here how to 
modify the equations when working in redshift space:

\be
P(\vk)\longrightarrow P(\vk)(1+\beta\mu^2)^2
\ee
(in the power spectrum only, not the bispectrum below);
\be
\langle\delta_{g\kvone}\delta_{g\kvtwo}\delta_{g\vkthree}\rangle
\longrightarrow {\cal G}(P_g,\beta,c_1,c_2,\kvone,\kvtwo,\vkthree) \;.
\ee

And finally the covariance matrix $C_{\alpha\beta}$ obtained 
with this prescription needs to be filtered:

\be
 C_{\nu_1 \nu_2}\longrightarrow \frac{C_{\nu_1\nu_2}}{\sqrt{\prod_{i=1}^6
{(1+k_i^2\sigma^2\mu_i^2/2)}}} \;,
\label{covar}
\ee
where the index $i$ runs over the six vectors that form the two triangles 
$\nu_1$ and $\nu_2$.   

\section{Tests on N-body simulations}

We have tested the model (equations \ref{bisps} \& \ref{covar}) by performing a
likelihood analysis of the bispectrum on an unbiased redshift-space catalogue
created from an N-body simulation provided by the Hydra consortium (Couchman et
al. 1995). This is a CDM simulation with $\Omega_0=1$, 
$\sigma_8=0.64$ and $\Gamma=0.25$; these parameters have been 
chosen to match closely the present day galaxy power spectrum. It is clear from
equations (\ref{bisp}), (\ref{G}) and (\ref{bisps}) that in a realistic
application we need to know (or have an accurate fit for) the real space galaxy
power spectrum, because it is this quantity that is required in the model and,
knowing it, subsequently we can attempt to evaluate $\sigma$. 

Therefore, as a first step, we need to know if we are able to reconstruct 
the real-space galaxy power spectrum.  The problem is greatly complicated 
by the fact that the method is applied on individual subvolumes, rather that
on the overall volume. Even assuming that the data will be  good enough to 
enable us to reconstruct the real-space power spectrum for the whole volume
of the survey, some care is needed in extracting then the real space 
spectrum for each subsample. A subsequent paper will focus on this issue, 
in the present paper we shall assume that we have an accurate enough fit
for the real space power spectrum. In Appendix B we give some limits 
on the accuracy needed.

\subsection{Limit of validity of the small scale redshift distortion model}

As mentioned in Section 2.4, the value $\sigma$ used in the model is only
 an ``effective'' velocity dispersion, so it can actually be seen as a 
parameter whose value is fixed by the condition that equation (6), when 
applied to  
the real space Fourier modes, gives the observed redshift space power
 spectrum.   
The limit of validity of the model, i.e. the maximum value of
 $\alpha=k^2\mu^2$, can be set as follows.

Since we can assume, as seen in Section 2.4, that the real-space power 
spectrum is known, a likelihood analysis of the redshift space Fourier 
modes should give the two parameters $\beta$ and $\sigma$.

The probability distributions of the real and imaginary parts of 
$\delta_{\vk}$ are Gaussian in linear theory, but the second order 
correction induces skewness.
However in this regime the contribution from the second order correction is
relatively small, and in any case, the Central Limit Theorem ensures that the
average of a set of modes from a region in $k-$space will
tend to a gaussian for a large number of modes, and the likelihood for the set
is then equivalent to a product of individual gaussians, since 
homogeneity alone ensures that the modes are uncorrelated.
Therefore, the likelihood tends to a product of Gaussians with zero means and 
dispersion $\sqrt{P^s(k,\mu)/2}$, where 
\be 
P^s(k,\mu)=P(k)(1+\beta\mu^2)^2D(k\sigma\mu)^2 \;.
\ee

The combined likelihood for $\beta$ and $\sigma$ 
\ba
& &\!\!\!\!\! \!\!\!\!\!\! {\cal L}(\beta,\sigma)= \nn
& &\!\!\!\!\!\!\!\!\!\!\!\! \frac{1}{(2\pi)^{M/2}\prod_{\nu}\left(
\frac{1}{2}P^s(k,\mu)\right)}\;\; e^{-\frac{1}{2}\sum_{\nu}
\frac{\{{\rm Re}\;\delta_{\vk}\}^2+\{{\rm Im}\;\delta_{\vk}\}^2}{\frac{1}{2}P^s(k,\mu)}} 
\;. \nn
\ea
The parameter 
$\beta$ in the simulation is known, but also in a realistic application 
will be known through an independent method (see the Introduction), therefore 
we can say that the  modelling for the small scale velocity dispersion 
effect breaks down where the likelihood fails to recover the ``true'' 
value for $\beta$.   
On large scales the model is expected to be valid for any value of the 
factor $\alpha=k^2\mu^2$, but  on smaller scales $\alpha$ must be constrained. 
We find no constraint on $\alpha$ up to $k=0.70$ then  $\alpha < 0.3 $ 
up to $k=0.9$.  For higher $k$ it becomes difficult to have a constraint 
on both $\beta$ and $\sigma$  for any value of $\alpha$. This feature 
is a direct consequence of an effect noted also by Bromley, Warren \& Zurek 
(1997): 
at short wavelengths the $\mu$ dependence is cancelling. However for
 $k>0.9$  $\sigma$ can still be recovered with less than 5 per cent uncertainty 
keeping $\alpha \leq 0.3 $ and constraining $\beta$ to assume the value 
previously determined.

We notice a slight increase of the value for $\sigma$ with $k$. This 
would suggest that the exponential model is not exact, but, as we
already pointed out, $\sigma$ can be treated as an empirical parameter.

\subsection{Obtaining the effective velocity dispersion in the subvolumes}

In the individual subvolumes the power spectrum becomes noisier, 
and the large-scale modes are not well sampled. Assuming that the 
local real space power spectrum is known, we can determine the local 
$\sigma$, because the local value for  $\beta$ can be extrapolated 
from the global one as follows:
if $N^s_{SV}$ is the number of particles in the subvolume in 
redshift space and $N_{\rm tot}$ is the number of particles in the whole 
volume of the sample 
\be
\beta_{{\rm local}}=\beta_{{\rm 
global}}\left(\frac{N^s_{SV}}{N_{{\rm tot}}}\right)^{0.6} \;.
\ee

There is a subtlety to point out here: in principle we would need to 
use the number of particles in the real space subvolume, but this is 
not known. However if the subsamples are not too small it should not 
make too much difference due also to the degeneracy that arises on 
those scales between $\beta$ and $\sigma$. Moreover an error of 20 per cent 
in the mean density (that is the biggest fluctuation we observe in 
splitting our simulation of $100 h^{-1}$ Mpc side in subsamples of $50 h^{-1}$
Mpc side) 
leads to an error of 5 per cent on $\beta_{\rm local}$ that is of the same order 
of the uncertainty on $\sigma$ obtained from the likelihood. 

Since we noticed a slight increase on the value for $\sigma$ with $k$,
the value for $\sigma$ has been obtained independently for different 
intervals of $k$.

We performed a likelihood analysis as described above, with  $\beta$ 
fixed, to recover $\sigma$ for each subvolume, in bands of $k$ limited by
the following values: 0.3, 0.7, 0.85, 1.0, 1.2, 1.3, with 
an additional band $1.7-2.6$.

The reason for this choice of intervals will be evident in the next section.

\section{Likelihood analysis}

It is evident from (\ref{bisps}) that a set of triangles 
in $k$-space allows the measurement of the bias parameters $b_1$ 
and $b_2$ through a likelihood analysis. It is however necessary to 
consider triangles of different shapes in order to lift 
a new degeneracy between $b_1$ and $b_2$ since a set of triangles  
of the same shape gives  almost degenerate information on $b_1$ 
and $b_2$ while  for the equilateral triangles the degeneracy is complete.
The optimal triangle configurations are not easy to determine, but it 
is particularly effective to consider, along with equilateral triangles, a 
configuration consisting of a repeated vector and one of twice the 
amplitude and opposite direction (referred to as `degenerate').
By ensuring that any $\vk$ appears in only one triangle of either shape 
(and $-\vk$ does not appear elsewhere either) the data are uncorrelated 
(MVH97): the covariance matrix is diagonal and the likelihoods for 
different shapes can be multiplied. This simplifies the likelihood analysis 
considerably.

\begin{figure*}
\begin{center}
\setlength{\unitlength}{1mm}

\begin{picture}(90,50)
\vspace{4cm}
\includegraphics{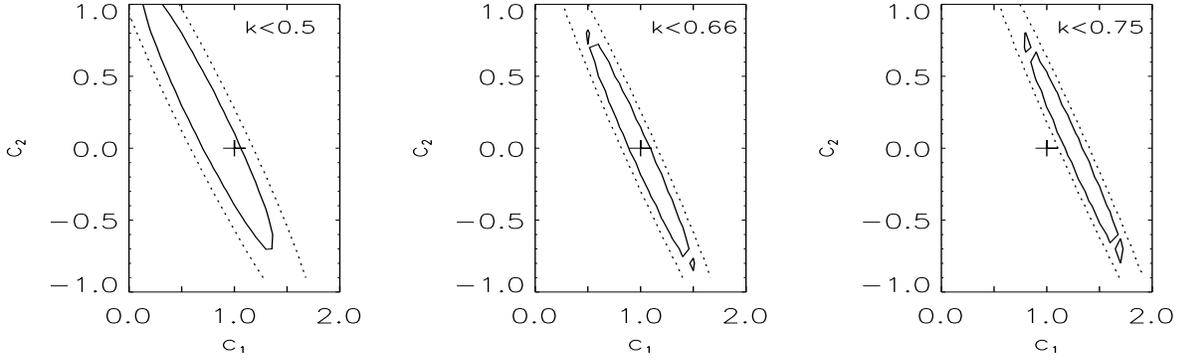}
\end{picture}
\end{center}
\label{c1est}
\caption{The real space joint likelihood for $c_1$ and $c_2$  as a function
of the cutoff wavenumber employed, for a low-density CDM simulation with shape 
parameter $\Gamma=0.25$ and $\sigma_8=1.06$. The dotted and solid
line contours contain 90 and 63 percent of the {\it a posteriori} probability
$P(c_1,c_2 \mid DATA)$ assuming uniform prior.
 The wavenumber corresponding to 
$\Delta^2(k)=1$ is $k=0.17$.  Second-order perturbation theory works
well for the degenerate configuration up to $k_{\rm short} \simeq 0.66$; this
corresponds to $\chi^3 \simeq 100$. This figure is relative to the final
output ($\sigma_8=1.06$) of the low-$\Omega_0$ simulation.}
\end{figure*}

\subsection{Choice of $k$-vector limits}

\subsubsection{$\Omega_0=1$ Universe}
Due to the smearing effect of small scale velocities the range of 
validity of 2OPT is expected to be  slightly extended.
This is exactly what we observe. For equilateral configurations for 
example, from Fig. 2 it is clear that the redshift-space bispectrum for 
equilateral triangles is well described by our model up to $k\simeq 0.7-0.9$. 

In real space the $k$-vector range we used for the likelihood 
analysis of the bispectrum was:
$0.3\leq k<0.55$ for equilateral configurations, and $0.55\leq k_{\rm short}< 1.1$ 
for degenerate.
In the redshift-space likelihood analysis for the bispectrum the range 
of $k$-vector used changes: for equilateral configurations $0.3 \leq k < 0.85$
 and, for degenerate configuration, $0.85\leq k_{\rm short}< 1.3$.  

\subsubsection{Low-$\Omega_0$ Universe}
In real-space the second-order bispectrum is only weakly dependent on
$\Omega_0$ (e.g. MVH97), in redshift-space there is an additional 
dependence on  
$\Omega_0$, but this is only through the parameter $\beta$ that can be measured (Heavens et
al. in preparation).
Although the remaining dependence on $\Omega_0$ is very weak, in a real survey we need to decide {\it a priori} where to place 
the upper cutoff in wavenumber.   Since, in principle, this will be dependent
on $\Omega_0$, we investigate a low-density numerical simulation to
see how robust the cutoff wavenumber is.
The simulation, again from the Hydra Consortium, is for a CDM model with
parameters 
$\Omega_0$=0.3, $\Lambda=0$, $\Gamma$=0.25, and a 
box-size of 100 $h^{-1}$ Mpc. 
Fig. 3 shows how the real space likelihood 
contour for $c_1$ and $c_2$ varies with the cutoff wavenumber 
for this simulation.
This final output has $\sigma_8$=1.06, which makes it unsuitable for our
redshift-space analysis. In fact the real space analysis recovers the true
value for the bias within the errors ($c_1=0.9\pm 0.5$, $c_2=0.1\pm 0.7$) but
the first mildly non-linear wavelengths are larger than the box and the 
redshift-space distortions are too non-linear for the wavelengths we need 
to use. We have therefore analysed 
two earlier epochs, with $\sigma_8=0.64$ (when $\Omega_0=0.5$) and 
$\sigma_8=0.81$ (when $\Omega_0=0.41$).

For the redshift-space analysis in these cases, since there is a strong 
covariance between $\beta$ and 
$\sigma$, it was not always possible for all the $k$-vector bands to 
follow the procedure outlined in 
Section 3 to set the values for the velocity dispersion parameter 
$\sigma$ 
and the angle $\alpha$. When it was not possible to close the likelihood
contours for $\beta$ and $\sigma$, we assumed $\beta$ was known to 
set the value for $\sigma$ and we used for $\alpha$ the same constraint
as in the other cases. 
 We also  find that the method for the redshift space analysis recovers the
true value of the bias parameter within the errors. The errors are still big
(100 per cent), but the non-linear scale is
also quite big ($k < 0.3$) and we can reduce the 
errors by analyzing subvolumes (MVH97). A rough 
estimate for the number of volumes that
one can obtain from the SDSS (about a thousand if linear theory
breaks down at $k_{\rm nl} \simeq 0.1$, about eight thousand if $k_{\rm nl} \simeq
0.3$) suggests that an accuracy of few percent or better should be achieved.

\begin{figure*}
\begin{center}
\setlength{\unitlength}{1mm}

\begin{picture}(90,100)
\includegraphics{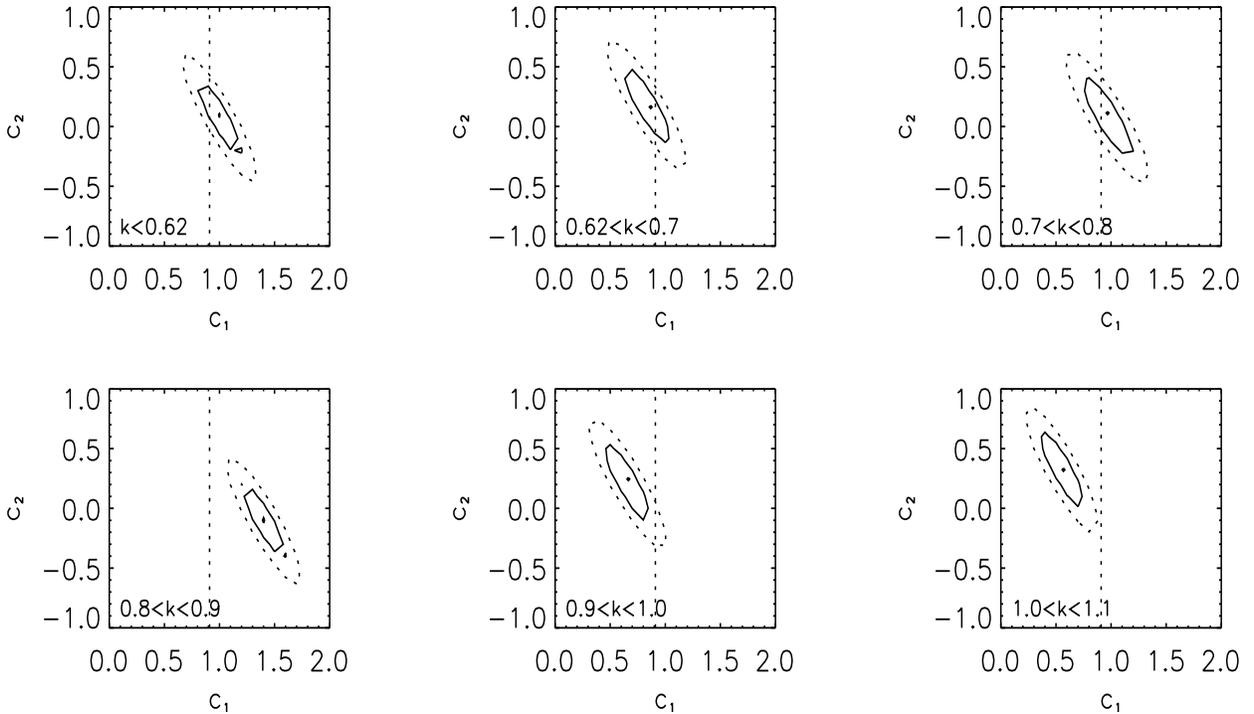}
\end{picture}
\end{center}
\label{c1estbias}
\caption{The real space joint likelihood for $c_1$ and $c_2$  for the biased
catalogue.The dotted and solid 
line contours contain 90 and 63 percent of the {\it a posteriori}
probability. In each
panel $k$-vectors of the
degenerate triangle configurations
belong to a different shell in $k$ space. For the equilateral configuration
the $k$-vectors limits are $0.3<k<0.5$ and for the degenerate the six shells
are limited by: 0.62, 0.7, 0.8, 0.9, 1.0, 1.1.
The first interval has been chosen in order to have about the same number of
data as in the following interval and a signal of comparable strength.
 Second-order perturbation theory works
well for the degenerate configuration up to $k_{\rm short} \simeq 0.8$ this
corresponds to $\chi^3 \simeq 100$.  The vertical dashed line shows the
inverse of the effective bias of the power spectrum, $\sqrt{P_g/P}$ at 
$k=0.8$.}
\end{figure*}

\subsubsection{General case}
The range of validity of second order perturbation theory depends on the model
of the Universe, so we therefore need to be able to determine, in a consistent
manner, where 2OPT breaks down prior to estimating $\Omega_0$. For all the
simulations we analyzed the breakdown occurs where the real-space normalized
bispectrum  $\chi^3=(2/\pi^2)k^6B$ (analogous to the quantity $\Delta^2$ for
the power spectrum) is $\chi^3 \simeq 10$ for equilateral triangles, and
$\simeq 100$ for degenerate triangles. This real-space quantity can be
estimated, assuming, as we do, that $\beta$ and the small-scale velocity
dispersion $\sigma$ can be estimated from the redshift-space power spectrum.
For the two earlier epochs of the low-$\Omega_0$ simulation we also changed
$\sigma_8$ by reinterpreting the box size (e.g. Mann, Peacock \& Heavens 1997):
also in these cases the breakdown occurs for the same values of the normalized 
bispectrum $\chi^3$. 

\subsection{Biased catalogue}

In order to create a biased catalogue, we applied the standard
friends-of-friends algorithm to the same $\Omega_0=1$ N-body simulation used in
MVH97 and in Section 2.1. 
We adopted a linking parameter equal to 15 per cent of the mean
interparticle distance, i.e. approximately $0.12 h^{-1}$ Mpc. The biased
catalogue is defined as the list of all the groups of particles with at least 
three 
members; the corresponding number density is $5.6\times 10^{-2} h^3$
Mpc$^{-3}$. Although this density is more realistic than that of the unbiased
catalogue ($2.1 ~h^3$ Mpc$^{-3}$) used in MVH97 and in the redshift-space
analysis, this is still about six times larger than 
the number density that future galaxy
surveys will have, but with this choice shot noise does not dominate the
signal in the regime where 2OPT should hold, allowing 
a determination of the breakdown of 2OPT.

This biasing scheme is not necessarily equivalent to the high peaks biasing
scheme nor can necessarily be analytically expressed as in 
equation (\ref{taylorexp}).
However for the purpose of recovering $\Omega_0$ from the $\beta$ parameter
what is necessary is the ``effective bias'' $b_{\rm eff}$ defined as:
$P_g=b_{\rm eff}^2P$. In the biased catalogue $b_{\rm eff}=1.1$ on scale where
2OPT holds.

Also in presence of biasing the breakdown of 2OPT can easily be detected.
Fig. 4 shows how the joint likelihood contours vary for $k$-vectors of
the degenerate triangle configuration belonging to different independent
shells in $k$-space. It is clear that 2OPT breaks down at $k\simeq 0.8$.
When perturbation theory breaks down the likelihood does not drift in a
precise direction as it happens for the unbiased case. This can be due to the
superposition of the non-linear clustering (that drifts the likelihood towards
high $c_1$) and the biasing acting in the opposite direction. 
For the biased case the $k$-vector limits are for the
equilateral configuration $0.3 < k < 0.5$ and for  degenerate configuration
$0.5<k_{\rm short}<0.8$. The breakdown occurs where the normalized bispectrum
$\chi^3=(2/\pi^2)k^6B$ is $\chi^3 \simeq10 $ for equilateral triangles and
$\chi^3 \simeq 100$ for degenerate triangles.        

\begin{figure}
\begin{center}
\setlength{\unitlength}{1mm}
\begin{picture}(90,70)
\includegraphics{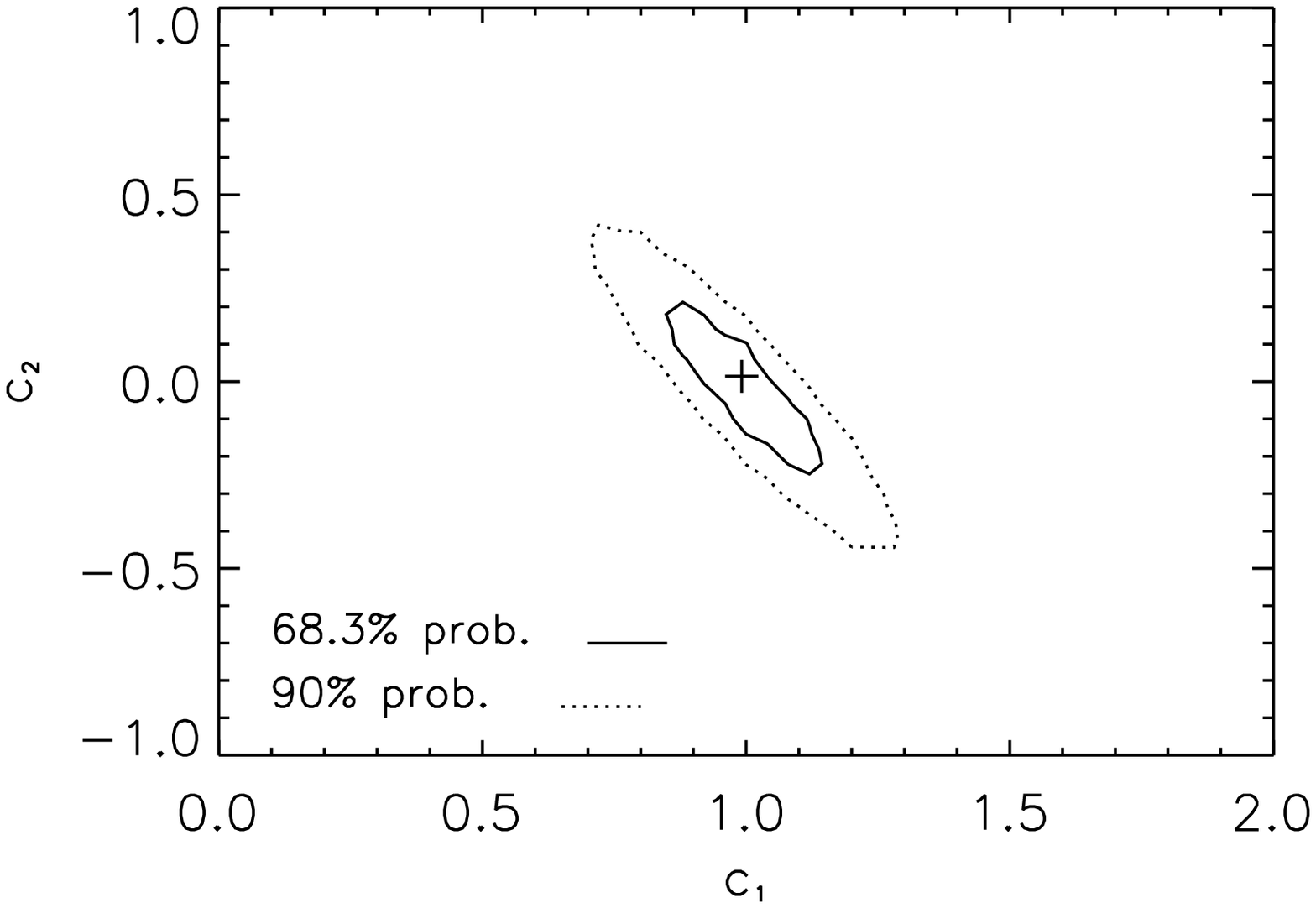}
\end{picture}
\end{center}
\label{lik}
\caption{Joint likelihood of $c_1=1/b_1$ and $c_2=b_2/b_1^2$, for 
a CDM N-body simulation ($\Omega_0=1$, $\sigma_8=0.64$, $\Gamma=0.25$, see text
for further details), but in redshift space. 
Contours contain 68.3 and 90 per cent of the {\it a posteriori} 
probability $P(c_1,c_2\mid DATA)$ assuming uniform priors for $c_1$ and $c_2$.
} 
\end{figure}
\begin{figure}
\begin{center}
\setlength{\unitlength}{1mm}
\begin{picture}(90,70)
\includegraphics{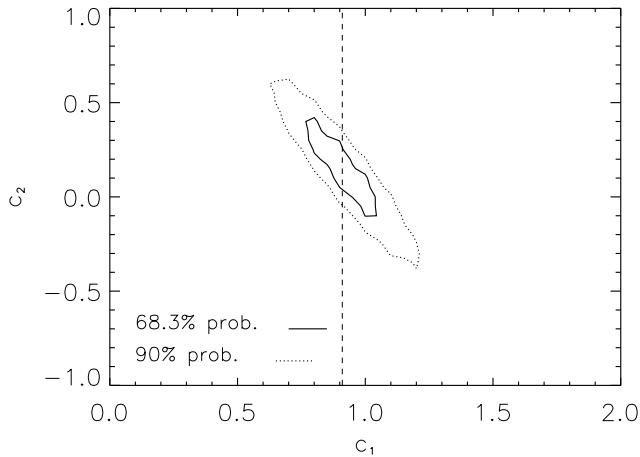}
\end{picture}
\end{center}
\label{likbias}
\caption{Joint likelihood of $c_1=1/b_1$ and $c_2=b_2/b_1^2$, for 
the real space N-body simulation (see text for details), but biased. The effective bias
on scales where 2OPT holds is $b_{\rm eff}$=1.1 ($1/b_{\rm eff}=0.91$) as
indicated by the dashed line. 
Contours contain 68.3 and 90 per cent of the {\it a posteriori} 
probability.} 
\end{figure}

We also checked the behaviour of the 2OPT validity range for a linearly biased
catalogue. The catalogue has been created by applying equation
(\ref{taylorexp}) with
$b_1=1.4$, $b_i=0$ $(i\geq 2)$, to the
density contrast field $\delta(\vec{x})$ of the $\Omega_0=1$, unbiased
simulation. In this case 2OPT breaks down for the same $k$-vectors values as
in the unbiased case therefore where $\chi^3$ is larger than the values
previously determined. However, unless there is anti-bias, the $\chi^3$ limit
sets a conservative $k$-vector cutoff.
\begin{figure}
\begin{center}
\setlength{\unitlength}{1mm}
\begin{picture}(90,70)
\includegraphics{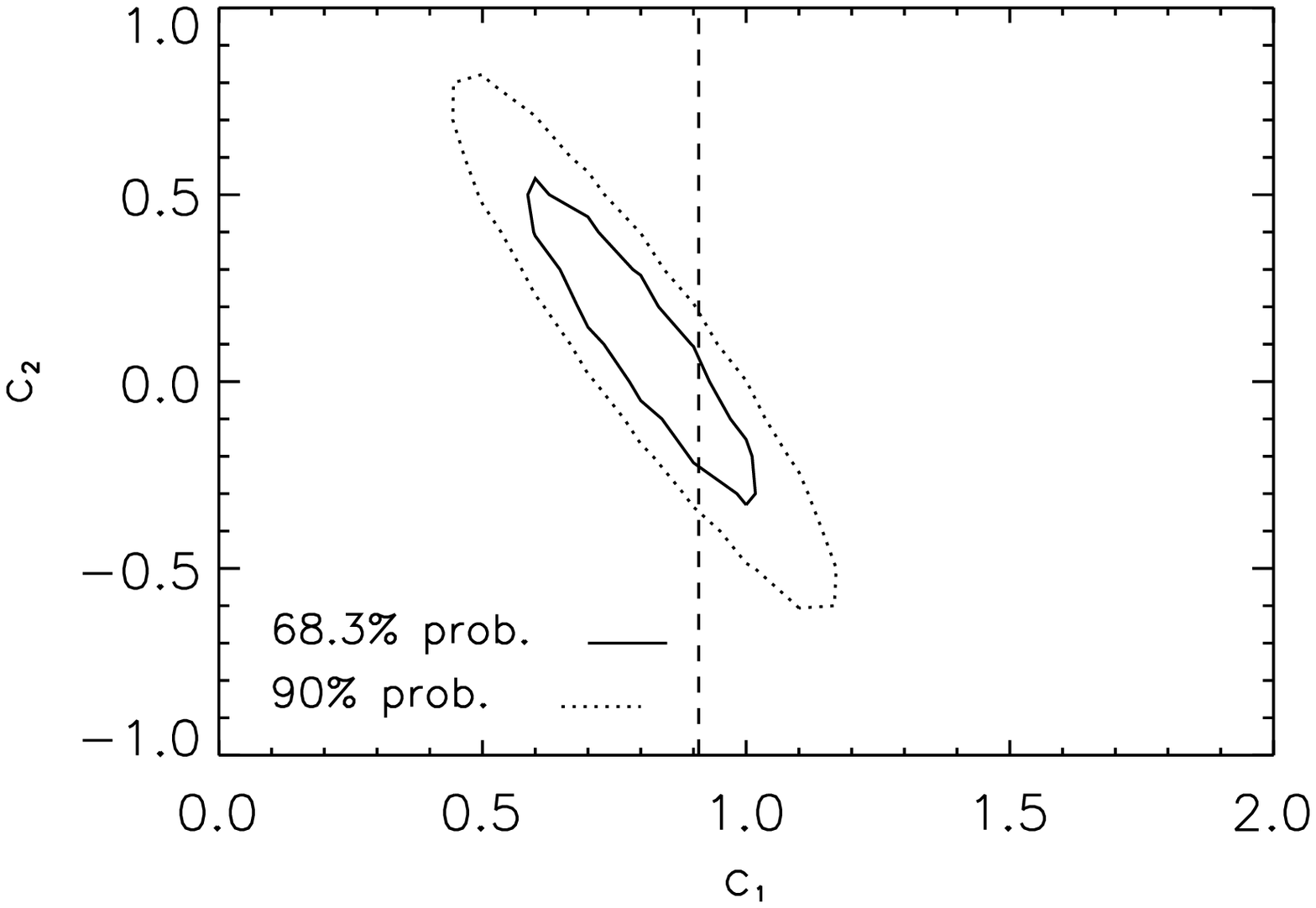}
\end{picture}
\end{center}
\label{likbiasred}
\caption{Joint likelihood of $c_1=1/b_1$ and $c_2=b_2/b_1^2$, for 
the biased simulation in redshift space. The effective bias
on scales where 2OPT holds is $b_{\rm eff}$=1.1 ($1/b_{\rm eff}=0.91$) as
indicated by the dashed line. 
Contours contain 68.3 and 90 per cent of the {\it a posteriori} 
probability.} 
\end{figure}

Finally we performed the likelihood analysis on the biased simulation with
$b_{eff}=1.1$ in redshift space.  Also in this more realistic case the
analysis recovers the true value for the bias parameter within the errors 
(see Fig. 7). In this case the errors in the determination of the bias
parameter are slightly bigger; this is due to the fact that the determination
of the velocity dispersion is more noisy. In fact the adopted biasing scheme 
considerably increases the shot noise, and this propagates in a noisy
determination of the velocity dispersion especially if the volume of the
simulation is limited to a box of 100 h$^{-1}$ Mpc side.  When applying the
method to a real survey the volume available will be much larger, allowing a
more precise determination of the velocity dispersion.
   
\subsection{Results}

For the flat Universe simulation in redshift space the joint likelihood for
$c_1$ and $c_2$ for the limits on $k$ set as described above, is shown in
Fig. 5.  The true value ($c_1=1/b_1$=1, $c_2=b_2/b_1^2$=0) is remarkably
recovered within the errors: $1.02\pm 0.15$ for $c_1$ and $-0.05 \pm 0.22$ for
$c_2$.

The joint likelihood computed in the biased simulation for $c_1$ and
$c_2$ for the limits on $k$ set as in Section 4.2 is shown in 
Fig. 6 and Fig. 7, in real and
redshift space respectively. Marginalising over $c_2$, the true value for the
effective bias $b_{\rm eff}=1.1$, $c_{\rm eff}=1/b_{\rm eff}$ is well
recovered through $c_1$: we obtain $c_1=0.93 \pm 0.15$, $c_2=0.15\pm0.25$ for
the real space case and $c_1=0.90 \pm 0.22$, $c_2=0.15\pm0.35$ for the
redshift space case.

\section{Discussion}

In Matarrese at al. (1997) we presented  the first steps towards the goal 
of using the bispectrum to measure the linear bias parameter $b$ from an 
ideal real-space galaxy distribution. The aim of this project is ultimately to 
combine measurements of $b$ with estimates of $\beta = \Omega_0^{0.6}/b$ 
from linear redshift distortion studies to get an estimate of the 
density parameter $\Omega_0$. In this paper, we have mainly tackled the problem
of redshift-space distortions,  where peculiar velocities distort the 
three-dimensional map in redshift surveys.  We have shown how the combination
of second-order perturbation theory and an incoherent velocity dispersion
model for virialised structures can successfully be used to estimate the
bias parameter.   This was not a trivial issue, since the signal for bias
comes largely from the mildly non-linear regime, which is significantly 
affected by non-perturbative considerations.  

We have also analyzed a biased simulation in real and redshift space: 
the bispectrum method
successfully recovers the value for the bias parameter not only if  we adopt a
linear bias scheme, but also in a more realistic case where groups of dark
matter particles are identified and each group counts as a galaxy.

To turn this into a practical application, one needs to be able to
determine, in a consistent manner, where second-order perturbation theory 
breaks down.  For the simulations we have analysed,  the breakdown occurs
when the real-space normalised bispectrum $\chi^3$ 
(analogous to the quantity $\Delta^2$ for the power spectrum) is $\chi^3 
\simeq 10$ for equilateral triangles, and $100$ for degenerate triangles.
This real-space quantity can be estimated, assuming, as we do, that $\beta$
and the small-scale velocity dispersion $\sigma$ can be estimated from
the redshift-space power spectrum.

The practical limitations to the accuracy of the determination of $b$ will
be the size of the redshift survey, and the density of objects.  Our 
suggestion for practical implementation is to divide the survey into a
number of subsamples, within each of which we apply the distant-observer 
approximation and make an estimate of the bias parameter (marginalising
over the second-order term $c_2$).  The size of the subsamples is 
determined by the non-linear scale, $k\simeq 0.3$ if $b\simeq 1$, 
restricting the subsample sizes to about 20 $h^{-1}$ Mpc or larger.
Since the signal comes from the weakly non-linear regime, the estimates
may be assumed to be essentially uncorrelated.
For the most distant subsamples, shot noise will prevent any useful information
being obtained.  This will restrict the depth to around $z=0.2$ for the 
forthcoming Anglo-Australian 2dF and SDSS. 
Nevertheless this gives a very large number (about 1800 and 
about 8000 respectively) of useful subvolumes (however, if the linear 
theory breaks down at $k_{\rm nl}\simeq 0.1$ the number of volume 
that one can obtain from the SDSS drops to about
a thousand). Combining this with the 
estimated error on $b$ from a single subvolume of about $50$ $h^{-1}$ Mpc 
suggests that 
errors of about 5 per cent or less should be achievable. The statistical error 
on $\beta$ from these surveys should be comparable or smaller 
(e.g. Ballinger et al. 1996; but see also Hatton \& Cole 
1998, Bromley et al. 1997), so the error on $\Omega_0$ will be 
mainly determined by the error on $b$ and should be around few percent.

\section*{Acknowledgments}

LV acknowledges the support of TMR grant.  Computations were made using 
STARLINK facilities. The simulations were obtained from the data bank 
of cosmological N-body simulations provided by the Hydra consortium 
(http://coho.astro.uwo.ca/pub/data.html) and produced using the 
Hydra N-body code (Couchman et al. 1995).  We are particularly grateful
to Peter Thomas for making earlier epochs of the simulations available.
AFH and LV thank the Dipartimento di Astronomia in Padova for hospitality.
SM and LM thank the University of Edinburgh for hospitality. 
Thanks to Francesco Lucchin for useful discussions, and to John Peacock
for comments on the manuscript.

\appendix

\section{Expression for the N-point galaxy spectra in redshift space}
Following the mathematical method outlined in Section 3 of MVH97 it is possible
to calculate the galaxy N-point spectra in redshift space. The expression for
the N-point spectra includes the large scale effect to second order in
perturbation theory (equation \ref {bisp}) and the small scale effect due to the
velocity dispersion of the galaxies with velocity distribution modelled by an
exponential (equation \ref{damping}).  

The expression for the external source ${\cal J}_k^d$ is still given by:
\be
{\cal J}_{\vec{k}}^d(\vec{x})=-iN \left\{ \exp \left[ \frac{i}{N}
\sum_{m=1}^N s_m e^{-i\vec{k}_m \cdot \vec{x}} \widetilde{W}(\vec{k}_m)
\right] -1 \right\},
\label{sourced}
\ee
but now the Fourier transform of the smoothing function 
($\widetilde{W}$) includes the effect of the filter: 
\be
\widetilde{W}(\vec{k}_m)=\frac{1}{\sqrt{1+k_m^2\sigma^2\mu^2/2}} \;.
\ee
The Ansatz for the generating functional is still the same: assuming 
Gaussian initial conditions and allowing for a quasi-linear evolution 
in 2OPT approximation all the irreducible correlation functions  $\xi^n_{\rm 
conn.}$ of order  $n > 3$ are negligible. But the difference is 
that in the expression for the generating functional

\be
\begin{array}{c}
\\
{\cal Z}[{\cal J}_k] =\\
\exp \left[ i \int d^3x {\cal J}_k(\vec{x})-\frac{1}{2}\int d^3x d^3 
x^{\prime}   
{\cal J}_k(\vec{x}){\cal J}_k(\vec{x}^{\prime} ) \xi^{(2)red}_{\rm conn.} 
(\vec{x},\vec{x}^{\prime} ) 
\right.\\
\left. - \frac{i}{6} \int d^3x d^3x^{\prime} d^3 x ^{\prime \prime} 
{\cal J}_k(\vec{x}) 
{\cal J}_k(\vec{x}^{\prime} ) {\cal J}_k(\vec{x}^{\prime \prime} ) 
\xi ^{(3)red}_{\rm conn.} (\vec{x},
\vec{x}^{\prime} ,\vec{x}^{\prime \prime} ) \right] \;,
\end{array}
\ee
now $\xi^{(2)red}_{\rm conn.}$ and $\xi ^{(3)red}_{\rm conn.}$ includes 
the redshift space effects that yield equations (\ref{bisp}) to (\ref{K2}) 
(Heavens et al., in preparation). 

With these modifications the generating functional approach outlined in 
MVH97 allows calculation of $N$-point spectra and their covariance properties  
directly in redshift-space taking also into account non-linearities 
to second order in perturbation theory.
The advantages are manifold since non-linearities affect the statistics of 
the density field even on quite large scales and, at the same time, there 
is the necessity to push the analysis to smaller scales to improve the 
statistics.  

The general prescription turns out to be 
that for a given configuration of $\{\vk_{i}\}_{i=1\ldots N}$
the $N$-point spectrum  $\langle \delta_{\vk_{1}}\ldots\delta_{\vk_{N}}\rangle$
 is first calculated consistently with 2OPT for the density field and the 
redshift distortion, and then has to be `filtered' with $N$ damping 
factors as in equation
(\ref{damping}) each one relative to one $\delta_{\vk_{i}}$. 

 The galaxy bispectrum in redshift space for a given triangle configuration
is a special case where $N=3$ and it is given in equation (\ref{bisps}).

\section{Constraints on the number of subsamples}

In MVH97 we showed that the signal to noise of the bias estimate can be 
improved by splitting the volume in subsamples. This is certainly true 
if the errors on the power spectra are negligible, but it may not be true 
if the subsamples  are too small and the spectra become noisy or if the 
real space power reconstruction procedure yields the power spectrum with 
uncertainties.

However it is possible to calculate the entity of the error on the bias 
parameter due to uncertainty in the power spectrum as follows.
We ignore shot noise for simplicity and we refer to equilateral triangles.
Since the likelihood adds information inversely weighted by variance, the 
error on $c_1$ $\sigma_{c_1}$ is given by
\be
\frac{1}{\sigma_{c_1}^2}=\sum_{shells}{\frac{1}{\sigma_{c_1}^2(shell)}} \;,
\ee 
$\sigma_{c_1}(shell)$ being the contribution to the  error on $c_1$ from a 
thin shell of radius $k$ in $k$-space.

This can be expressed in terms of the error on the power spectrum.
Let $P(k)$ be the average power in the shell in $k$-space; $P(k)$ is constant
in the shell and has a variance $\sigma^2_{P(k)}$.
Therefore
\be
\sigma_{c_1}^2(shell)=\left( \frac{2 \sum_\nu}{\sum_\nu P(k)}\right)^2
\sigma_{P(k)}^2=\frac{4 \sigma^2_{P(k)}}{P(k)^2} \;,
\ee
where $\nu$ labels the triangles.

Summing the contributions from all the shells,
\be 
\sigma_{c_1}^2=\frac{1}{\sum_{shells}\frac{P^2}{4\sigma_P^2}}\ ,
\ee
and passing to the continuum limit, under the hypothesis that $P$ is 
affected only by statistical error, we obtain 
\be
\sigma^2_{c_1}=\frac{4}{\frac{V 4 \pi}{2(2\pi)^3}\int_{k_{\rm 
min}}^{k_{\rm max}}k^2 dk}=\frac{\gamma^2}{V} \;,
\ee
where $V$ is the volume of the sample in units of $h^{-3}$ Mpc$^3$. 
If $k_{\rm max}$ is the same as in real space ($k_{\rm max}$ 
previously found was 0.55),
we have $\sigma_{c_1}\simeq53/\sqrt{V}$. As we have seen, the redshift-space
 $k_{\rm max}$ can be pushed towards higher $k$, reducing the error on $c_1$: 
if $k_{\rm max}=0.85$ as we found here, then  $\sigma_{c_1}\simeq28/\sqrt{V}$
    
Therefore splitting the volume in $M$ subvolumes is still effective if 
\be 
\frac{\sigma_{c_1}^2(B)}{M} \gg \frac{ \gamma M}{V_{SV}}= \frac{ \gamma}
{V_{\rm tot}} \;,
\ee
where $\sigma_{c_1}(B)$ is the error on $c_1$ due to the variance in the 
bispectrum, $V_{SV}=V_{\rm tot}/M$ is the volume of the subsamples. 
This is a constraint on the number of subsample to make, once the 
properties of the selection function of the  survey and the range of 
validity of 2OPT are known.   

As long as the errors on the reconstructed real-space power spectrum are 
of the same order of its statistical errors, and if still, as we found 
in the simulation, $\sigma_{c_1}^2(B)\simeq 0.5$ and the optimal size 
for the subvolumes is $\simeq 20h^{-1}$ Mpc (MVH97), then the use of the 
subvolumes is still effective both for SDSS and 2dF surveys. 
Any additional uncertainty on the power spectrum will alter the right-hand
side of equation (B5) 
placing a stronger constraint on the number of subvolumes and/or on their size.

\end{document}